\documentclass{ws-procs9x6-cpt19}
\begin{document}

\def\al{\alpha}
\def\be{\beta}
\def\ga{\gamma}
\def\de{\delta}
\def\ep{\epsilon}
\def\ve{\varepsilon}
\def\ze{\zeta}
\def\et{\eta}
\def\th{\theta}
\def\vt{\vartheta}
\def\io{\iota}
\def\ka{\kappa}
\def\la{\lambda}
\def\vpi{\varpi}
\def\rh{\rho}
\def\vr{\varrho}
\def\si{\sigma}
\def\vs{\varsigma}
\def\ta{\tau}
\def\up{\upsilon}
\def\ph{\phi}
\def\vp{\varphi}
\def\ch{\chi}
\def\ps{\psi}
\def\om{\omega}
\def\Ga{\Gamma}
\def\De{\Delta}
\def\Th{\Theta}
\def\La{\Lambda}
\def\Si{\Sigma}
\def\Up{\Upsilon}
\def\Ph{\Phi}
\def\Ps{\Psi}
\def\Om{\Omega}
\def\pt#1{\phantom{#1}}
\def\prt{\partial}
\def\cl{{\cal L}}
\def\half{\tfrac12}
\def\qrt{\tfrac14}

\newcommand{\beq}{\begin{equation}}
\newcommand{\eeq}{\end{equation}}
\newcommand{\bea}{\begin{eqnarray}}
\newcommand{\eea}{\end{eqnarray}}
\newcommand{\refeq}[1]{(\ref{#1})}

\def\mn{{\mu\nu}}
\def\rs{{\rh\si}}

\def\syjm#1#2{{}_{#1}Y_{#2}}

\def\q{q}
\def\qHat{\widehat{\q}}
\def\qd#1{{\q}^{(#1)}}

\def\s{s}
\def\sHat{\widehat{\s}}
\def\sd#1{{\s}^{(#1)}}

\def\k{k}
\def\kHat{\widehat{\k}}
\def\kd#1{{\k}^{(#1)}}

\def\kjm#1#2#3{k^{(#1)}_{(#2)#3}}
\def\kI{\kjm{d}{I}{jm}}
\def\kE{\kjm{d}{E}{jm}}
\def\kB{\kjm{d}{B}{jm}}
\def\kV{\kjm{d}{V}{jm}}
\def\kIdjm#1#2{\kjm{#1}{I}{#2}}
\def\kEdjm#1#2{\kjm{#1}{E}{#2}}
\def\kBdjm#1#2{\kjm{#1}{B}{#2}}
\def\kVdjm#1#2{\kjm{#1}{V}{#2}}

\title{Nonminimal Lorentz Violation in Linearized Gravity}

\author{Matthew Mewes}

\address{Physics Department, California Polytechnic State University\\
San Luis Obispo, CA 93407, USA}

\begin{abstract}
  This contribution to the CPT'19 meeting
  provides a brief overview
  of recent theoretical studies
  of nonminimal Lorentz violation
  in linearized gravity.
  Signatures in gravitational waves
  from coalescing compact binaries
  are discussed.
\end{abstract}

\bodymatter

\phantom{}\vskip10pt\noindent
The Standard-Model Extension (SME)
is a general framework for studies
of arbitrary realistic violations of
Lorentz and CPT invariance.
The SME has provided a theoretical base
for hundreds of searches
for Lorentz and CPT violations
in particles and in gravity.\cite{tables}
The leading-order violations in the particle sectors
of the SME were written down
more than two decades ago,\cite{sme}
followed by the leading-order violations
in gravity.\cite{smegrav}
These violations modify
the Standard Model of particle physics and General Relativity.
Together they give the so-called minimal Standard-Model Extension (mSME).

A Lorentz-violating term in
the SME action takes the form
of a conventional tensor operator
contracted with a tensor coefficient
for Lorentz violation:
\begin{equation}
\de S = \int d^4x\
({\rm coefficient~tensor})
\cdot({\rm tensor~operator}) \ .
\end{equation}
The tensor coefficients for Lorentz violation
act as Lorentz-violating background fields.
Each violation can be classified according to
the mass dimension $d$ of the conventional operator
in natural units with $\hbar=c=1$.
The mSME contains the violations of renormalizable
dimensions $d=3,4$.
Nonminimal violations are those with $d\geq 5$.
Nonminimal extensions have been constructed
for a number of sectors of the SME,
including gauge-invariant electromagnetism,\cite{kmph}
neutrinos,\cite{kmnu}
free Dirac fermions,\cite{kmferm}
quantum electrodynamics,\cite{qed}
General relativity,\cite{cubic,bkxsr}
and linearized gravity.\cite{bkxsr,ktcr,kmgw1,kmnewt,kmgw2,gw19}

The extension for linearized gravity
includes all possible modifications
to the usual linearized Einstein-Hilbert
action that are quadratic in the metric
fluctuation $h_\mn = g_\mn - \et_\mn$.
Each unconventional term takes the form\cite{kmgw1}
\beq
\de S
= \int d^4x\,
\qrt {\mathcal K}^{(d)\mn\rs\al_1\ldots\al_{d-2}}
h_\mn \prt_{\al_1}\ldots \prt_{\al_{d-2}} h_\rs \ ,
\eeq
where ${\mathcal K}^{(d)\mn\rs\al_1\ldots\al_{d-2}}$
are the coefficients for Lorentz violation.
The coefficient tensors can be split
into irreducible pieces with unique symmetries,
giving fourteen different classes
of Lorentz violation.\cite{kmgw2}
Three of these classes yield modifications
that are invariant under
the usual gauge transformation
$h_\mn \rightarrow h_\mn + \prt_{(\mu} \xi_{\nu)}$.
Restricting attention to the gauge-invariant violations,
the Lorentz-violating parts can be written as\cite{kmgw1}
\beq
S_\text{LV} =
\int d^4x\, \tfrac14 h_\mn 
(\sHat{}^{\mu\rh\nu\si} 
+ \qHat{}^{\mu\rh\nu\si} 
+ \kHat{}^{\mu\nu\rh\si})
h_\rs \ ,
\eeq
where the three operators
\bea
\sHat{}^{\mu\rh\nu\si} &=&
\sum \sd{d}{}^{\mu\rh\al_1\nu\si\al_2\ldots\al_{d-2}}
\prt_{\al_1}\ldots \prt_{\al_{d-2}} \ ,
\notag \\
\qHat{}^{\mu\rh\nu\si} &=&
\sum \qd{d}{}^{\mu\rh\al_1\nu\al_2\si\al_3\ldots\al_{d-2}}
\prt_{\al_1}\ldots \prt_{\al_{d-2}} \ ,
\notag \\
\kHat{}^{\mu\nu\rh\si} &=&
\sum \kd{d}{}^{\mu\al_1\nu\al_2\rh\al_3\si\al_4\ldots\al_{d-2}}
\prt_{\al_1}\ldots \prt_{\al_{d-2}}
\label{sqk}
\eea
contain the three types of gauge-invariant violations.
The $\s$- and $\k$-type violations are CPT even,
while $\q$-type violations break CPT invariance.
The sums in Eq.\ \refeq{sqk} are over
even $d\geq4$ for $\s$-type violations,
odd $d\geq5$ for $\q$-type, and
even $d\geq6$ for $\k$-type.

The gauge invariant limit provides
a simple framework for
studies of Lorentz violation in gravity,
including studies of short-range gravity,\cite{bkxsr,kmnewt}
gravitational \v Cerenkov radiation,\cite{ktcr}
and gravitational waves.\cite{kmgw1,kmgw2,gw19}
For gravitational waves,
the violations give a modified phase velocity of the form\cite{kmgw1}
\beq
v = 1 - \vs^0 \pm \sqrt{|\vs_{(+4)}|^2 + |\vs_{(0)}|^2} \ .
\label{vel}
\eeq
The effects of Lorentz violation are controlled
by the frequency- and direction-dependent functions
\bea
\vs^0
&=& \sum_{djm} \om^{d-4} \syjm{0}{jm}(-\hat v)\, \kI \ ,\notag \\
\vs_{(\pm4)} 
&=& \sum_{djm} \om^{d-4} \syjm{\pm4}{jm}(-\hat v)\, (\kE \pm i \kB) \ , \notag \\
\vs_{(0)}
&=& \sum_{djm} \om^{d-4} \syjm{0}{jm}(-\hat v)\, \kV \ ,
\label{sigma}
\eea
where $\om$ is the angular frequency.
Spin-weighted spherical harmonics $\syjm{s}{jm}$
are used to characterize
the dependence on the direction of propagation $\hat v$.
The spherical coefficients for Lorentz violation 
$\kI$, $\kV$, $\kE$ and $\kB$
are complicated linear combinations of the underlying
tensor coefficients in Eq.\ \refeq{sqk}.

\begin{figure}[t]
\begin{center}
\includegraphics[width=3.3in]{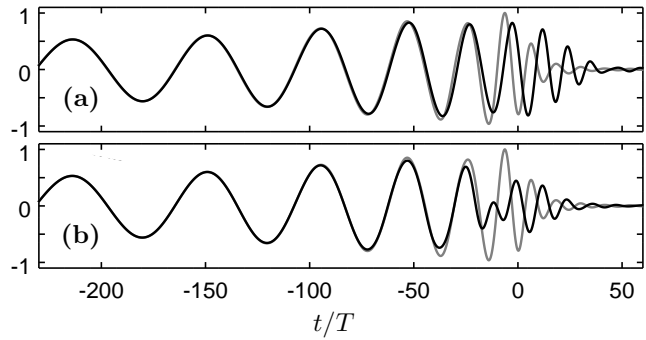}
\end{center}
\caption{
  Simulated noise-free detector strain signals
  from the merger of two equal-mass black holes for
  (a) nonbirefringent dispersion with $\vs^{(6)0} = 20T^3/\ta$
  and
  (b) birefringence with $\vs_{(0)}^{(5)} = 10T^2/\ta$.
  The effective propagation time $\ta$ accounts for
  the redshift of the frequency during propagation.
  We also define
  the characteristic merger time scale $T = G_N (1+z)M$
  in terms of the Newton's constant $G_N$
  and the merger's redshift $z$
  and total mass $M$.
  The plots show
  the Lorentz-violating cases in black
  and the Lorentz-invariant limit in gray.\cite{gw19}}
\label{figs}
\end{figure}

The Lorentz violation associated with $\kI$ coefficients
produce a frequency-dependent velocity,
leading to direction-dependent dispersion
in gravitational waves.
An example of the effects of this type of violation
in a binary merger is shown in the top plot of Fig.\ \ref{figs}.
In this example,
$d=6$ violations lead to
a shift in phase velocity that is proportional $\om^2$.
This particular shift causes
the higher-frequency components of the wave
to travel slower than lower-frequency parts.
Early in the merger, when lower frequencies dominate,
the effects of dispersion are insignificant.
Higher frequencies dominate
at later times,
where the signal experiences
a delayed arrival,
deforming the tail end of the waveform.

The violations associated with the
$\kV$, $\kE$ and $\kB$ coefficients
yield two distinct propagating solutions.
The two solutions have different polarizations
that are determined by the
$\vs_{(+4)}$,  $\vs_{(-4)}$, and $\vs_{(0)}$ combinations.\cite{gw19}
Each solution propagates at a different speed,
corresponding to the two signs in Eq.\ \refeq{vel}.
This gives rise to birefringence (in addition to dispersion).
A general wave is a superposition
of the two solutions,
which results in a net polarization
that evolves as the wave propagates,
yielding a key signature of birefringence.
The effects depend on frequency,
so each frequency experiences a different
change in polarization.

An example of the effects
of birefringence are illustrated
in the bottom plot of Fig.\ \ref{figs}.
In this example,
$d=5$ birefringent Lorentz violations
produce a simple rotation of
the polarization of the wave.
The rotation angle grows with $\om$,
so higher frequencies experience a greater change.
The example assumes that the gravitational wave
is linearly polarized
and that the arms of the detector
are aligned so that the strain signal
is maximized in the Lorentz-invariant limit.
At later stages in the merger,
the higher frequencies produce
a greater rotation of the polarization.
This affects the relative alignment
of the arms of the detector and the wave's polarization,
decreasing the response of the detector
and distorting the signal.

\section*{Acknowledgments}
This work was supported in part 
by the United States National Science Foundation 
under grant number 1819412.

\end{document}